\documentclass[aps,prd,showpacs,twocolumn]{revtex4}
\usepackage{amsmath}
\usepackage{amsfonts}
\usepackage{amssymb}
\usepackage{bm}

\begin{document}

\title{Lorentz-violating effects in the Bose-Einstein condensation of an
ideal bosonic gas}
\author{Rodolfo Casana and Kleber A. T. da Silva}
\affiliation{Departamento de F\'{\i}sica, Universidade Federal do Maranh\~{a}o (UFMA),\\
Campus Universit\'{a}rio do Bacanga, S\~{a}o Lu\'{\i}s, MA, 65080-805,
Brazil.}

\begin{abstract}
We have studied the effects of Lorentz-violation in the Bose-Einstein
condensation (BEC) of an ideal boson gas, by assessing both the
nonrelativistic and ultrarelativistic limits. Our model describes a massive
complex scalar field coupled to a CPT-even and Lorentz-violating background.
We first analyze the nonrelativistic case, at this level by using
experimental data, we obtain upper-bounds for some LIV parameters. In the
sequel, we have constructed the partition function for the relativistic
ideal boson gas which to be able of a consistent description requires the
imposition of severe restrictions on some LIV coefficients. In both cases,
we have demonstrated that the LIV contributions are contained in an overall
factor, which multiplies almost all thermodynamical properties. An exception
is the fraction of the condensed particles.
\end{abstract}

\maketitle

\section{Introduction}

The CPT- and Lorentz-symmetry violations have been intensively investigated
in the latest years. A strong motivation to study the CPT- and
Lorentz-symmetry breaking is the necessity to get some information about
underlying physics at Planck scale where the Lorentz symmetry may be broken
due to quantum gravity effects, possibility opened up in early 90's \cite%
{Samuel1,Samuel2,Samuel3,Samuel4,Samuel5,Samuel6,Samuel7}. Another reason is
the need of examining the limits of validity of the CPT theorem and the
Lorentz symmetry, based on the search for small deviations from scenarios
characterized by CPT and Lorentz symmetry exactness. This line of
investigation is conducted mainly in two contexts, one within the framework
of the Standard Model Extension (SME) \cite%
{Samuel1,Samuel2,Samuel3,Samuel4,Samuel5,Samuel6,Samuel7,Colladay1,Colladay2}
and another into the framework of the Planck scale modified dispersion
relations \cite{Jacobson1,Jacobson2,Jacobson3,Jacobson4}. The SME
incorporates terms governing the effects of the spontaneous symmetry
breaking of the CPT- and Lorentz-invariance in all sectors of the Standard
Model of the fundamental interactions and particles. The main researches are
devoted to the study of LIV effects in classical and quantum electrodynamics
with the objectives to establish strong upper limits over the parameters
ruling the CPT- and Lorentz-violating effects. In this sense, a set of many
investigations and diverse experimental setups, based on distinct theories
and effects, have been proposed to constrain the Lorentz-violation
parameters leading to the upper limits presented in Ref. \cite{LIV-DATA}.

The study of LIV effects in statistical physics into the context of the SME
has been initiated in Ref.\cite{Colladay-stat}, based on the maximum entropy
approach. There, it was then considered a general nonrelativistic
Hamiltonian, containing the Lorentz-violating terms coming from the SME
fermion sector \cite%
{Fermion1,Fermion2,Fermion3,Fermion4,Fermion5,Fermion6,Fermion7,Fermion8}.
It was shown that the usual laws of thermodynamics remain unaffected and
that the relevant corrections appear at the form of rotationally invariant
functions of the LIV parameters. The theoretical framework developed in Ref.
\cite{Colladay-stat} was used to analyze the influence of Lorentz violation
on Bose-Einstein condensation in Ref. \cite{Colladay-bec}\textbf{. }It was
shown the Lorentz-violating terms can change the shape and the phase of the
ground-state condensate produced by means of trapping techniques. An
alternative study of BEC in the LIV framework was recently performed in Ref.
\cite{mpla-1} via the use of deformed dispersion relation in statistical
physics. Moreover, the LIV effects in other thermodynamical systems, as the
electromagnetic sector of the SME, have been examined in Refs. \cite%
{Casana1a,Casana1b,Casana2,Casana3} starting from a
finite-temperature-field-theory approach. Specifically, it was studied the
influence of the Lorentz-violating CPT-odd and CPT-even terms on the black
body radiation and the anisotropies induced in the angular energy density
distribution.

In this work, we discuss some Lorentz-violating effects on a bosonic system,
described by a complex scalar field, able to support the Bose-Einstein
condensation \cite{BEC-books1,BEC-books2,BEC-books3}. It is important to
remark that the Bose-Einstein condensation can open up a new route for
searching for small deviations of Lorentz symmetry if refined and accurate
experimental techniques are used. Therefore, Bose-Einstein condensation
could provide a new set of laboratory tests relevant for restricting
Lorentz-violation parameters even more. Our aim is to study the effects of
the Lorentz-violation in the Bose-Einstein condensation of an ideal boson
gas in both the nonrelativistic and ultrarelativistic limits.

\section{A CPT-even and Lorentz-violating model for the complex scalar field}

The simplest Lorentz-invariance violating Lagrangian for the complex scalar
field in (1+3)-dimensions is
\begin{equation}
\mathcal{L}=\partial _{\mu }\phi ^{\ast }\partial ^{\mu }\phi +\lambda ^{\mu
\nu }\partial _{\mu }\phi ^{\ast }\partial _{\nu }\phi -m^{2}\phi ^{\ast
}\phi -U(|\phi |),  \label{lr-01}
\end{equation}%
where $\lambda ^{\mu \nu }$ is a dimensionless symmetric tensor ruling the
CPT-even and Lorentz-invariance violating contributions, $U(|\phi |)$ is an
self-interaction potential. The model described by (\ref{lr-01}) can be
interpreted an Abelian $O\left( 2\right) -$scalar model, a particular case
of the one studied in Ref. \cite{Altschulx1}. The second term in Lagrangian (%
\ref{lr-01}) was already used to study Lorentz-violating effects on
topological defects generated by scalar fields in (1+1) dimensions \cite%
{dionisio1}. A similar term has been also adopted to study the influence of
Lorentz violation on the relativistic version of acoustic black holes
generated in an Abelian Higgs model \cite{Brito1,Brito2}. Further, in the
context of the aether-like models, the model (\ref{lr-01}) can be considered
as an extension of those obtained in Ref. \cite{aether}.

It is worthwhile to observe that a CPT-odd LIV term $i\kappa ^{\mu }\left(
\phi ^{\ast }\partial _{\mu }\phi -\phi \partial _{\mu }\phi ^{\ast }\right)
$ could be added to the Lagrangian (\ref{lr-01}) but it can be eliminated by
an appropriate canonical field redefinition
\begin{equation}
\phi \rightarrow e^{i\hat{\kappa}\cdot x}\varphi ~,~\ \phi ^{\ast
}\rightarrow e^{-i\hat{\kappa}\cdot x}\varphi ^{\ast },
\end{equation}%
with $\hat{\kappa}^{\mu }$ chosen as $\hat{\kappa}^{\mu }=\left( g^{\mu \nu
}+\lambda ^{\mu \nu }\right) ^{-1}\kappa _{\nu }$. Note that the inverse of
the expression in parentheses does exist because the Lorentz-violating
parameter $\lambda ^{\mu \nu }$ is small compared to 1. By expressing the
Lagrangian (\ref{lr-01}) in terms of the new field $\varphi $, we get
\begin{equation}
\mathcal{L}\rightarrow \partial _{\mu }\varphi ^{\ast }\partial \varphi
+\lambda ^{\mu \nu }\partial _{\mu }\varphi ^{\ast }\partial _{\nu }\varphi
-\left( m^{2}+\hat{\kappa}_{\mu }\kappa ^{\mu }\right) \varphi ^{\ast
}\varphi -U(|\varphi |),
\end{equation}%
which no longer exhibits the CPT-odd term depending on $\kappa ^{\mu },$
whereas the mass term acquires a small correction. Then, if the\ term $%
i\kappa ^{\mu }\left( \phi ^{\ast }\partial _{\mu }\phi -\phi \partial _{\mu
}\phi ^{\ast }\right) $ can be removed with such a canonical transformation,
it is not a true LIV term and the $\kappa ^{\mu }$ vector cannot be
measurable \cite{Altschulx2}. Therefore, from now on we will only consider
the CPT-even and Lorentz-violating term in the lagrangian density for the
complex scalar field given in (\ref{lr-01}).

It is known that in the absence of a potential $U(|\phi |)$ in (\ref{lr-01}%
), the Lorentz-violation can be absorbed by a coordinate transformation and
a field rescaling \cite{Tadson}, however the free model can be considered as
a tree level of a more complete theory. In this context, it is worth
considering the implications of the Lorentz-violating in the Bose-Einstein
condensation of an ideal bosonic gas both in the nonrelativistic limit as in
ultrarelativistic. Thus, in the following we consider the model (\ref{lr-01}%
) with null potential.

\section{Lorentz-violating effects in a nonrelativistic ideal boson gas}

For applications of the model (\ref{lr-01}) in low energy situations, we
compute its associated nonrelativistic lagrangian density,
\begin{equation}
\mathcal{L}^{\prime }=i\gamma \psi ^{\ast }\partial _{t}\psi -\psi ^{\ast }%
\mathcal{H}_{C}^{\prime }\psi ,  \label{lagnr-01}
\end{equation}%
where we have define the coefficient $\gamma $,
\begin{equation}
\gamma =1+\lambda _{00},  \label{gama}
\end{equation}%
being a positive-definite quantity, and $H_{C}^{\prime }$ is the canonical
hamiltonian density given by%
\begin{equation}
\mathcal{H}_{C}^{\prime }=-\frac{1}{2m}\nabla ^{2}+\frac{1}{2m}\lambda
_{jk}\partial _{j}\partial _{k},  \label{Hamiltonian}
\end{equation}%
where we have neglected the terms linear in derivatives because they do not
contribute to the system energy \cite{Colladay-bec}.

Hence, the modified Schr\"{o}dinger equation generated by the
nonrelativistic Lagrangian density is
\begin{equation}
-i\gamma \partial _{t}\psi +\mathcal{H}_{C}^{\prime }\psi =0.  \label{eqsh}
\end{equation}%
The coefficient $\gamma $\ in the first term plays the role of the Planck's
constant ($h=1$ in natural units) modified by the Lorentz-violating
coefficient $\lambda _{00}$. Such a circumstance allows to use Eq. (\ref%
{gama}) to find an upper-bound to $\lambda _{00}$. For such a purpose we use
the relative uncertainty $\left( \Delta h/h\right) $, thus we attain
\begin{equation}
\left\vert \lambda _{00}\right\vert \leq 3.6\times 10^{-8},  \label{loo1}
\end{equation}%
where we have used the best value for the ratio, $\Delta h/h=3.6\times
10^{-8}$,\ provided by CODATA\ 2006 \cite{CODATA2006}.

If we perform the following field rescaling in (\ref{lagnr-01}) : $\psi
\rightarrow \gamma ^{-1/2}\psi $ and $\psi ^{\ast }\rightarrow \gamma
^{-1/2}\psi ^{\ast }$, it reads as%
\begin{equation}
\mathcal{L}^{\prime \prime }=i\psi ^{\ast }\partial _{t}\psi -\frac{1}{%
\gamma }\psi ^{\ast }\mathcal{H}_{C}^{\prime }\psi ,  \label{lagnr-1}
\end{equation}%
such that the modified Schr\"{o}dinger equation would be read as%
\begin{equation}
-i\partial _{t}\psi +\frac{1}{\gamma }\mathcal{H}_{C}^{\prime }\psi =0.
\label{eqsh-1}
\end{equation}%
By looking the second term, we can set an upper limit\ for $\lambda _{00}$
by using the relative uncertainty of $\left( \hbar ^{2}/2m\right) ~$that can
be obtained using the respective relative uncertainties of $\left(
h/m\right) $ and $\left( h\right) $. For example, for the $^{\text{87}}$Rb
atom we get
\begin{equation}
\left\vert \lambda _{00}\right\vert \leq 4.9\times 10^{-8}  \label{loo2}
\end{equation}%
which is compatible with the upper-bound obtained in Eq.(\ref{loo1}).

In the remaining of this section we use the Lagrangian density (\ref{lagnr-1}%
) to analyzed the Lorentz-violating effects in the BEC of a nonrelativistic
ideal boson gas.

The Lagrangian density (\ref{lagnr-1}) is invariant under the following
global field transformation: $\psi \rightarrow e^{-i\alpha }\psi ~,~~\psi
^{\ast }\rightarrow e^{i\alpha }\psi ^{\ast }$, whose conserved charge
density is $\psi ^{\ast }\psi $. Then, the partition function is defined by
\begin{equation}
Z\left( \beta \right) =\int \mathcal{D}\psi \mathcal{D}\psi ^{\ast }\mathcal{%
~}\exp \left\{ -\int_{\beta }dx~\psi ^{\ast }\mathbf{D}\psi \right\} ,
\label{z-bec0}
\end{equation}%
where the operator $\mathbf{D}$ is
\begin{equation}
\mathbf{D}=\partial _{\tau }-\frac{1}{2m\gamma }\nabla ^{2}+\frac{1}{%
2m\gamma }\lambda _{jk}\partial _{j}\partial _{k}-\mu ,  \label{z-bec1}
\end{equation}%
where $\mu $ is the chemical potential associated to the conserved charge
density $\psi ^{\ast }\psi $. The integration is performed over the fields
satisfying periodical boundary conditions in the $\tau $ variable: $\psi
\left( \tau ,\mathbf{x}\right) =\psi \left( \tau +\beta ,\mathbf{x}\right) ~$%
and$~\psi ^{\ast }\left( \tau ,\mathbf{x}\right) =\psi ^{\ast }\left( \tau
+\beta ,\mathbf{x}\right) $.

In absence of LIV interactions, the operator defined in Eq. (\ref{z-bec1})
is $\displaystyle\left( \partial _{\tau }-\frac{1}{2m}\nabla ^{2}-\mu
\right) ,$ whose zero-mode is intimately related with the existence of the
Bose-Einstein condensation \cite{shakel,superconductivity}. It allows to
guarantee that the BEC phenomenon occurs when $\mu \rightarrow 0^{-}$. The
value $\mu =0$ is the fundamental value for occurring BEC. Such condition is
similar to the superconductivity phase transition: it only happens when the
value of the resistivity is zero \cite{superconductivity}. It is clear that
the LIV term $\lambda _{jk}\partial _{j}\partial _{k}$ in Eq. (\ref{z-bec1})
does not modify the zero-mode condition so the BEC in a LIV framework also
occurs when $\mu \rightarrow 0^{-}$.

Therefore, by computing the functional integration (\ref{z-bec0}) in the
Fourier space, the partition function becomes
\begin{equation}
\ln Z(\beta )=-V\!\int \!\frac{d^{3}\mathbf{p}}{\left( 2\pi \right) ^{3}}%
\sum_{n}\ln \left[ i\beta \omega _{n}+\beta \left( \tilde{\epsilon}-\mu
\right) \right] ,  \label{nrz1}
\end{equation}%
with $\omega _{n}$\ being the bosonic Matsubara's frequencies, $\omega
_{n}=2\pi n/\beta ,~n=0,\pm 1,\pm 2,\ldots $ and, $\tilde{\epsilon}=$ $%
\tilde{\epsilon}\left( \mathbf{p}\right) $ is the particle's kinetic energy
with Lorentz-violating contributions,%
\begin{equation}
\tilde{\epsilon}\left( \mathbf{p}\right) =\frac{1}{2m\gamma }%
N_{jk}p_{j}p_{k},  \label{exx}
\end{equation}%
here we have defined a symmetric\ matrix $\mathbb{N=}\left[ N_{jk}\right] $
whose components are given by
\begin{equation}
N_{jk}=\delta _{jk}-\lambda _{jk},  \label{nn}
\end{equation}%
it will be positive-definite if the components$\ \lambda _{jk}$\ are
sufficiently small. By performing the summation in Eq.(\ref{nrz1}), we get%
\begin{equation}
\ln Z\left( \beta \right) =-V\int \frac{d^{3}\mathbf{p}}{\left( 2\pi \right)
^{3}}\ln \left[ 1-e^{-\beta \left( \tilde{\epsilon}-\mu \right) }\right] .
\label{nrz2}
\end{equation}%
Now, we make the following operations under the momentum integral: We first
perform a rotation $\mathbf{p}\rightarrow \mathbb{R}\mathbf{p}$, such that $%
\mathbb{R}$ diagonalizes the matrix $\mathbb{N}$, i. e., $\mathbb{R}^{T}%
\mathbb{NR}=\mathbb{D}$, where $\mathbb{D}$ is a diagonal matrix whose
elements are the eigenvalues of $\mathbb{N}$. Next, we make the following
rescaling $\mathbf{p}\rightarrow \gamma ^{1/2}\mathbb{D}^{-1/2}\mathbf{p}$.
Under such transformations, the particle energy reads%
\begin{equation}
\tilde{\epsilon}\left( \mathbf{p}\right) =\frac{1}{2m}\mathbf{p}%
^{2}=\epsilon \left( \mathbf{p}\right) ,  \label{eyy}
\end{equation}%
is the usual free particle's energy in absence of Lorentz violation. On the
other hand, the partition function (\ref{nrz2}) becomes
\begin{equation}
\ln Z(\beta )=\gamma ^{3/2}(\det \mathbb{N})^{-1/2}\ln Z^{\left( 0\right) },
\label{zzzx}
\end{equation}%
where $Z^{\left( 0\right) }(\beta )$\ is the partition function of a
nonrelativistic bosonic ideal gas%
\begin{equation}
\ln Z^{\left( 0\right) }=-V\int \!\!\frac{d^{3}\mathbf{p}}{\left( 2\pi
\right) ^{3}}\ln \left[ 1-e^{-\beta \left( \mathbf{\epsilon }-\mu \right) }%
\right] ,
\end{equation}%
with $\epsilon =\epsilon \left( \mathbf{p}\right) $\ given by Eq. (\ref{eyy}%
). Observe that in Eq. (\ref{zzzx}) \ the Lorentz-violating contributions
are contained in the overall factor $\gamma ^{3/2}\left( \det \mathbb{N}%
\right) ^{-1/2}$. It is clear that in absence of Lorentz-violation, i. e., $%
\gamma =1$ and $N_{ij}=\delta _{ij}$, we recuperate the usual partition
function of the nonrelativistic complex scalar field.

\subsection{The CPT-even and Lorentz-violating effects in the
nonrelativistic BEC}

The nonrelativistic particle density is given by
\begin{equation}
n=\frac{1}{2\pi ^{2}}\gamma ^{3/2}\left( \det \mathbb{N}\right)
^{-1/2}\int_{0}^{\infty }dp~\frac{p^{2}}{e^{\beta \left( \mathbf{\epsilon }%
-\mu \right) }-1}.  \label{eq51-1y}
\end{equation}%
Here, the chemical potential must be\ negative $\left( \mu <0\right) $, once
the particle density is non-negative. Note that Eq.(\ref{eq51-1y})\ is an
implicit formula for $\mu $ as a function of $\rho $ and $T$. For $T$ above
some critical temperature $T_{_{C}}$, one can always find one value of $\mu $%
\ for which Eq. (\ref{eq51-1y}) holds. If the particle density held fixed in
$n=\bar{n}$ and the temperature is lowered, $\mu $ $\rightarrow 0^{-}$, in
the region $T\geq T_{_{C}}$ one achieves
\begin{equation}
\bar{n}=\left( \frac{m}{2\pi \beta }\right) ^{3/2}\zeta \left( 3/2\right)
\gamma ^{3/2}\left( \det \mathbb{N}\right) ^{-1/2},  \label{ssx}
\end{equation}%
while the critical temperature is
\begin{equation}
T_{_{C}}=T_{0}\gamma ^{-1}\left( \det \mathbb{N}\right) ^{1/3},~~T_{0}=\frac{%
2\pi }{m}\left( \frac{\bar{n}}{\zeta \left( 3/2\right) }\right) ^{2/3},
\label{tc-nonrel0}
\end{equation}%
where $T_{0}$ is the BEC critical temperature in absence of LIV terms. By
written the critical temperature at first order in LIV coefficients, we get
\begin{equation}
T_{_{C}}=T_{0}\left[ 1-\lambda _{00}-\frac{1}{3}\text{tr}\left( \lambda
_{ij}\right) +\ldots \right] ,  \label{tc-nonrel}
\end{equation}%
the contribution of the terms $\lambda _{ij}$ was also observed in the
results of Refs.\cite{Colladay-stat,Colladay-bec}. However, the contribution
of $\lambda _{00}$ to the critical temperature is one of our results.

The expansion in (\ref{tc-nonrel}) can be used to establish an upper limit
for the parameter tr$\left( \lambda _{ij}\right) $ by using experimental
data for the relative uncertainty of the BEC temperature. Such a
temperature, $T_{0}$, is experimentally determined in the range 0.5--2$\mu $%
K \cite{superconductivity}. On the other hand, the most refined experiments
are able to detect temperature fluctuations of the order of $0.5\times
10^{-10}$K \cite{lowest}. Actually, such lower temperatures can be accessed
with laser cooling techniques. By considering temperature fluctuations of
the order of $10^{-11}$K as the experimental uncertainty in a BEC
temperature measurement, the relative uncertainty for BEC's temperature
would be $5\times 10^{-6}$--$2\times 10^{-5}$ allowing to establish the
following upper-bound for the LIV coefficients
\begin{equation}
\left\vert \lambda _{00}+\frac{1}{3}\text{tr}\left( \lambda _{ij}\right)
\right\vert <5\times 10^{-6}.  \label{loo3}
\end{equation}%
By considering the upper-bound for $\lambda _{00}$\ attained in Eqs. (\ref%
{loo1},\ref{loo2}) we can establish an estimative upper bound for $\lambda
_{ij}$\ so restrictive as%
\begin{equation}
\left\vert \text{tr}\left( \lambda _{ij}\right) \right\vert <1.5\times
10^{-5}  \label{loo4}
\end{equation}

At temperatures $T<T_{_{C}}$, the expression (\ref{ssx}) becomes an equation
for charge density $\bar{n}-n_{0}$ of the nonzero momentum\textbf{\ (}$%
\mathbf{p\neq 0)}$ states,
\begin{equation}
\bar{n}-n_{0}=\left( \frac{m}{2\pi \beta }\right) ^{3/2}\zeta \left(
3/2\right) \gamma ^{3/2}\left( \det \mathbb{N}\right) ^{-1/2}\,,
\end{equation}%
where $n_{0}$ is the density of the condensed particles, so \ the density in
the ground state ($\mathbf{p=0)}$ is
\begin{equation}
n_{0}=\bar{n}\left[ 1-\left( \frac{T}{T_{_{C}}}\right) ^{3/2}\right] .
\end{equation}%
This shows that the fraction of the condensate density is not modified by
the Lorentz-violation. Also, the condition for nonrelativistic BEC, $\bar{n}%
\ll m^{3},$ is maintained.

\section{The relativistic ideal boson gas in a CPT-even and
Lorentz-violating framework}

The model of Lagrangian (\ref{lr-01}) is invariant under the $U\left(
1\right) $ global symmetry, $\phi \rightarrow e^{-i\alpha }\phi \ $and $\phi
^{\ast }\rightarrow e^{i\alpha }\phi ^{\ast }$, where $\alpha $ is any real
constant. The conserved charge density, expressed in terms of the fields $%
\phi $ and $\phi ^{\ast }$ and their respective conjugate\ momenta $\pi
^{\ast }$ and $\pi ,$ is
\begin{equation}
j^{0}=i\left( \phi ^{\ast }\pi -\phi \pi ^{\ast }\right) ,\
\end{equation}%
which has the same canonical structure of the Lorentz invariant case. By
considering $U(|\phi |)=0$, the canonical Hamiltonian density is given by
\begin{eqnarray}
\mathcal{H}_{C} &=&\gamma ^{-1}\pi ^{\ast }\pi +\nabla \phi ^{\ast }\cdot
\nabla \phi -\lambda _{jk}\partial _{j}\phi ^{\ast }\partial _{k}\phi
+m^{2}\phi ^{\ast }\phi ,  \notag \\
&& \\
&&\hspace{-0.5cm}+\gamma ^{-1}\lambda _{0j}\left( \pi ^{\ast }\partial
_{j}\phi +\pi \partial _{j}\phi ^{\ast }\right) +\gamma ^{-1}\lambda
_{0j}\lambda _{0j}\partial _{j}\phi ^{\ast }\partial _{j}\phi ,  \notag
\end{eqnarray}%
is positive-definite for $\lambda ^{\mu \nu }$ sufficiently small. Thus, the
partition function is defined as%
\begin{eqnarray}
Z\left( \beta \right) &=&\int \mathcal{D}\phi \mathcal{D}\phi ^{\ast }%
\mathcal{D}\pi \mathcal{D}\pi ^{\ast } \\
&&\hspace{-0.75cm}\times \exp \left\{ \int_{\beta }dx\left[ ~i\pi ^{\ast
}\partial _{\tau }\phi +i\pi \partial _{\tau }\phi ^{\ast }-\mathcal{H}%
_{C}+\mu j^{0}\right] \right\} ,  \notag
\end{eqnarray}%
where $\mu $\ is the chemical potential. The functional integration is
performed over the fields satisfying periodic boundary conditions, $\phi
\left( \tau +\beta ,\mathbf{x}\right) =\phi \left( \tau ,\mathbf{x}\right) ,$
and $\phi ^{\ast }\left( \tau +\beta ,\mathbf{x}\right) =\phi ^{\ast }\left(
\tau ,\mathbf{x}\right) $. By performing the momentum integrations and some
integrations by parts in the sequel, the partition function takes the form%
\begin{equation}
Z\left( \beta \right) =\int \mathcal{D}\phi \mathcal{D}\phi ^{\ast }\exp
\left\{ -\int_{\beta }dx~\phi ^{\ast }\mathbf{D}\phi \right\} ,
\label{bec-21}
\end{equation}%
where%
\begin{equation}
\mathbf{D}=-\eta \left( \partial _{\tau }-\mu \right) ^{2}+2\lambda _{\tau
k}\left( \partial _{\tau }-\mu \right)\partial _{k} -N_{jk}\partial
_{j}\partial _{k}+m^{2},  \label{drel}
\end{equation}%
with $N_{jk}$ the matrix defined in (\ref{nn}), we also have made the
following definitions: $\eta =1-\lambda _{\tau \tau }>0$,$~~\lambda _{\tau
\tau }=-\lambda _{00}$,~and $\lambda _{\tau j}=-i\lambda _{0j}$. The
existence of the zero mode for the operator defined in Eq. (\ref{drel}) is
intimately related of the relativistic BEC phenomenon. So, in a
Lorentz-violating framework, it implies that when the chemical potential
attains the value $\left\vert \mu \right\vert =\eta ^{-1/2}m$ the
relativistic Bose-Einstein condensation begins happen. In absence of
Lorentz-violation, $\eta=1$, the relativistic BEC occurs when $\left\vert
\mu \right\vert =m$ \cite{weldon1,weldon2,kapusta}.

The functional integration in (\ref{bec-21}) is computed in the momentum
space, yielding
\begin{equation}
\ln Z\left( \beta \right) =-V\int \frac{d\mathbf{p}}{\left( 2\pi \right) ^{2}%
}\sum\limits_{n}\ln \left[ \beta ^{2}\mathbf{\tilde{D}}\left( n,\mathbf{p}%
\right) \right] ,  \label{ss1}
\end{equation}%
with
\begin{equation}
\mathbf{\tilde{D}}\left( n,\mathbf{p}\right) =\eta (\omega _{n}+i\mu
)^{2}-2\lambda _{\tau j}(\omega _{n}+i\mu )p_{j}+N_{jk}p_{j}p_{k}+m^{2},
\end{equation}%
where $\omega _{n}=\frac{2\pi n}{\beta },~n=0,\pm 1,\pm 2,...,$ are the
bosonic Matsubara's frequencies,

Performing the summation in (\ref{ss1}), the partition function becomes%
\begin{eqnarray}
\ln Z &=&\ln \bar{Z}^{\left( +\right) }+\ln \bar{Z}^{\left( -\right) },
\notag \\[-0.2cm]
&&  \label{ss1a} \\
\ln \bar{Z}^{\left( \pm \right) } &=&-V\int \frac{d^{3}\mathbf{p}}{\left(
2\pi \right) ^{3}}\ln \left( \frac{{}}{{}}1-\exp \left[ -\beta \left( \bar{%
\epsilon}\mp \bar{\mu}\right) \right] \right) ,  \notag
\end{eqnarray}%
we have introduced the following definitions
\begin{eqnarray}
\bar{\epsilon} &=&\sqrt{\eta ^{-1}\left( N_{jk}p_{j}p_{k}+m^{2}\right)
-\left( \eta ^{-1}\lambda _{\tau {j}}p_{j}\right) ^{2}},  \label{ss1ba} \\
&&  \notag \\
\bar{\mu} &=&\mu +i\eta ^{-1}\lambda _{\tau {j}}p_{j},  \label{ss1bb}
\end{eqnarray}%
From (\ref{ss1bb}) we observe that the chemical potential gains an imaginary
part that is momentum dependent, $\eta ^{-1}\lambda _{\tau j}p_{j}$, which
changes the bosonic character of the field, however,\ the possibility of
statistical transmutation does not happen in 3D physical systems. Therefore,
a consistent description of the relativistic ideal bosonic gas in a CPT-even
and Lorentz-violating framework implies that the coefficients $\lambda
_{\tau k}$ be null, \emph{i.e.},
\begin{equation}
\lambda _{\tau k}=0.  \label{ljk0}
\end{equation}%
Under such physical requirement, the partition function (\ref{ss1a}) becomes%
\begin{eqnarray}
\ln Z &=&\ln \tilde{Z}^{\left( +\right) }+\ln \tilde{Z}^{\left( -\right) },
\notag \\[-0.2cm]
&&  \label{rpfxx} \\
\ln \tilde{Z}^{\left( \pm \right) } &=&-V\int \frac{d^{3}\mathbf{p}}{\left(
2\pi \right) ^{3}}\ln \left( \frac{{}}{{}}1-\exp \left[ -\beta \left( \tilde{%
\epsilon}\mp \mu \right) \right] \right) ,  \notag
\end{eqnarray}%
where
\begin{equation}
\tilde{\epsilon}=\sqrt{\eta ^{-1}(N_{jk}p_{j}p_{k}+m^{2})}.
\end{equation}%
so the partition function describes a relativistic ideal boson gas composite
of charged particles in a Lorentz-violating framework. We now perform the
following operations under the momentum integral. First, we perform a
rotation $\mathbf{p}\rightarrow \mathbb{R}\mathbf{p,}$ such that $\mathbb{R}$
diagonalizes the matrix $\mathbb{N}$, i. e., $\mathbb{R}^{T}\mathbb{NR}=%
\mathbb{D}$, where $\mathbb{D}$ is a diagonal matrix whose elements are the
eigenvalues of $\mathbb{N}$. Finally, we make the rescaling $\mathbf{p}%
\rightarrow \eta ^{1/2}\mathbb{D}^{-1/2}\mathbf{p}$. Hence, the partition
function (\ref{rpfxx}) becomes
\begin{eqnarray}
\ln Z &=&\eta ^{3/2}\left( \det \mathbb{N}\right) ^{-1/2}\left[ \ln
Z^{\left( +\right) }+\ln Z^{\left( -\right) }\right] ,  \notag \\[-0.2cm]
&&  \label{zzzy} \\
\ln Z^{\left( \pm \right) } &=&-V\int \frac{d^{3}\mathbf{p}}{\left( 2\pi
\right) ^{3}}\ln \left[ 1-e^{-\beta \left( \omega \mp \mu \right) }\right] ,
\notag
\end{eqnarray}%
where we have defined
\begin{equation}
\omega =\sqrt{\mathbf{p}^{2}+\eta ^{-1}m^{2}}.
\end{equation}%
We observe that in Eq. (\ref{zzzy}) the LIV contributions are contained in
the factor $\eta ^{3/2}\left( \det \mathbb{N}\right) ^{-1/2}$. Also , both
integrals converge if $\left\vert \mu \right\vert \leq \eta ^{-1/2}m$ and
the ultrarelativistic BEC occurs when
\begin{equation}
\mu =\pm \eta ^{-1/2}m,
\end{equation}%
so the chemical potential for relativistic BEC condensation is modified by
Lorentz-violation, in total accordance with zero mode analysis. It is clear
to note that in absence of Lorentz-violation,\ $N_{ij}=\delta _{ij}$, $\eta
=1$, we recover the partition function for the relativistic complex scalar
field \cite{weldon1,weldon2,kapusta}.

\subsection{The CPT-even and Lorentz-violating contributions to the
ultrarelativistic BEC}

We follow Ref.\cite{weldon1,weldon2} to describe the ultrarelativistic BEC
in this CPT-even and Lorentz-violating framework. Thus, for $\left\vert \mu
\right\vert <M$\ the charge density is
\begin{eqnarray}
\rho &=&\eta ^{3/2}\left( \det \mathbb{N}\right) ^{-1/2}\left[ \tilde{\rho}%
^{\left( +\right) }+\tilde{\rho}^{\left( -\right) }\right]  \notag \\[-0.2cm]
&&  \label{eq51-1x} \\
\tilde{\rho}^{\left( \pm \right) } &=&\int \!\!\frac{d^{3}\mathbf{p}}{(2\pi
)^{3}}\frac{1}{e^{\beta (\omega \mp \mu )}-1}.  \notag
\end{eqnarray}%
This equation is really an implicit formula for $\mu $ as a function of $%
\rho $ and $T$. For $T$ above some critical temperature $T_{_{C}}$, one can
always find a value for $\mu $ such that Eq. (\ref{eq51-1x}) holds. If the
density $\rho =\bar{\rho}$ is maintained fixed and the temperature is
lowered, the chemical potential $\mu $ increases until the point $\left\vert
\mu \right\vert =\eta ^{-1/2}m$ is reached. Thus, in the region $T\geq
T_{_{C}}\gg \eta ^{-1/2}m$, we obtain
\begin{equation}
\left\vert \bar{\rho}\right\vert \approx \frac{1}{3}m\left( \det \mathbb{N}%
\right) ^{-1/2}\eta T^{2}~.  \label{eq51-1x1}
\end{equation}%
When $\left\vert \mu \right\vert =\eta ^{-1/2}m$ and the temperature is
lowered even further such\ that $T<T_{_{C}}$, the charge density is written
as
\begin{equation}
\bar{\rho}=\rho _{0}+\rho ^{\ast }(\beta ,\mu =\eta ^{-1/2}m)\,,
\label{eq51-2x}
\end{equation}%
where $\rho _{0}$ is a charge contribution from the condensate (the
zero--momentum mode) and the $\rho ^{\ast }(\beta ,\mu =\eta ^{-1/2}m)$ is
the thermal particle excitations (finite--momentum modes) which is given by
Eq. (\ref{eq51-1x}) with $\left\vert \mu \right\vert =\eta ^{-1/2}m$.

The critical temperature $T_{_{C}}$, in which the Bose--Einstein
condensation occurs, is reached when $\left\vert \mu \right\vert =\eta
^{-1/2}m$, and is determined implicitly by the equation $\bar{\rho}=\rho
^{\ast }(\beta _{_{C}},\mu =\eta ^{-1/2}m)$, so that%
\begin{equation}
T_{_{C}}=T_{0}\left( \det \mathbb{N}\right) ^{1/4}\eta ^{-1/2},~T_{0}=\left(
\frac{3\left\vert \bar{\rho}\right\vert }{m}\right) ^{1/2},  \label{trel}
\end{equation}%
where $T_{0}$ is the ultrarelativistic critical temperature \cite%
{weldon1,weldon2} in absence of the Lorentz-violation. By expanding at first
order in LIV parameters we obtain
\begin{equation}
T_{_{C}}=T_{0}\left[ 1+\frac{1}{2}\lambda _{\tau \tau }-\frac{1}{4}\text{tr}%
\left( \lambda _{ij}\right) +\ldots \right] .
\end{equation}

At temperatures $T<T_{_{C}}$, expression (\ref{eq51-2x}) is an equation for
the charge density $\bar{\rho}-\rho _{0}$ of the nonzero momentum ($p\neq 0)$
states,
\begin{equation}
\bar{\rho}-\rho _{0}=\frac{1}{3}m\left( \det \mathbb{N}\right) ^{-1/2}\eta
T^{2}\,,  \label{eq51-4}
\end{equation}%
so that the charge density in the ground state ($p=0)$ is
\begin{equation}
\rho _{0}=\bar{\rho}\left[ 1-\left( \frac{T}{T_{_{C}}}\right) ^{2}\right] \,.
\end{equation}%
Thus, differently from the critical temperature and the chemical potential,
the fraction of the condensate charge is not modified by the
Lorentz-violating terms. So, the necessary condition for an ideal Bose gas
of mass $m$ to undergo a\ Bose-Einstein condensation in LIV backgrounds at
ultrarelativistic temperature ($T_{_{C}}\gg \eta ^{-1/2}m$) is that $\bar{%
\rho}\gg M^{3}$, \textit{i.e.},
\begin{equation*}
\bar{\rho}\gg (1-\lambda _{\tau \tau })^{-3/2}m^{3}\sim \left( 1+\frac{3}{2}%
\lambda _{\tau \tau }\right) m^{3},
\end{equation*}%
hence such a condition is slightly modified by Lorentz-violation.

\section{Remarks and conclusions}

We have studied the Bose-Einstein condensation of an ideal Bose gas in a
Lorentz-violating framework in both limits, the nonrelativistic and the
ultrarelativistic ones. The model is described by a complex scalar field
containing a CPT-even and Lorentz-violating term. We first have studied the
nonrelativistic limit of the model which gains only contributions of the
parity-even LIV coefficients. The experimental data of Planck's constant and
rubidium mass allows to obtain consistent upper-bounds for $\lambda _{00}$
coefficient given by Eqs. (\ref{loo1}), (\ref{loo2}). A second set of
upper-bounds (see Eqs. (\ref{loo3}) and (\ref{loo4})) was obtained by using
the experimental data for BEC's temperature and measurements of lower
temperatures obtained via laser cooling techniques. It is clear that an
analysis of our model in more realistic situations such as considering
trapping potentials or particle interactions would give an enhancement to
our study. However, the description would not change substantially and the
upper-bounds for Lorentz-violating coefficients will remain almost the same.

In the relativistic case, a consistent definition of a partition function
describing properly the statistical behavior of charged bosons requires $%
\lambda _{\tau j}=0$ ($\lambda _{\tau j}$ is related to the parity-odd
coefficient $\lambda _{0j}$). In nonrelativistic case, it has been shown
that the partition function in LIV background is a power of the partition
function in absence of Lorentz-violation, such power contains the full
contribution of Lorentz-violation. This way, such power is an overall factor
which multiplies all thermodynamical properties. However, there are some
exceptions, as for example, the fraction of the condensed particles and the
value of the chemical potential where BEC happens are unaffected by
Lorentz-violation. In relativistic case, it has been shown that a part of
the LIV contributions are contained in an overall factor which multiplies
all thermodynamical properties. The factorization is not complete because
the chemical potential has Lorentz-violating contributions which are not
contained in the global factor. However, the fraction of the condensed
charged is unaltered by Lorentz-violation.

\begin{acknowledgments}
RC thanks to CNPq, CAPES and FAPEMA (Brazilian research agencies) by partial
financial support and KATS thanks to CAPES by full financial support. The
authors thank to M. M. Ferreira Jr. for reading the paper and helpful
comments and suggestions.
\end{acknowledgments}

\end{document}